# Translating database mathematical schemes into relational database software applications with *MatBase*

**Christian Mancas** CHRISTIAN.MANCAS@GMAIL.COM
*Ovidius University at Constanta, Romania.*
**Diana Christina Mancas** DIANA.CHRISTINA.MANCAS@GMAIL.COM
*Ovidius University at Constanta, Romania.*

**Corresponding Author:** Christian Mancas.

## Abstract

We present a pseudocode algorithm for translating our (Elementary) Mathematical Data Model schemes into relational ones and associated sets of non-relational constraints, used by *MatBase*, our intelligent data and knowledge base management system prototype. We prove that this algorithm is very fast, solid, complete, and optimal. We apply it to a Mathematical scheme modeling the genealogical trees subuniverse. We also provide examples of SQL and VBA code for enforcing some of its non-relational constraints, as well as guidelines to develop code for enforcing such constraints.

## 1. INTRODUCTION

Every subuniverse of discourse is governed by business rules. For example, stocks may not be negative, people may not live more than 140 years, countries have only one capital city, dynasties only one founder, etc. Databases (dbs) and their managing software applications should enforce all of them to guarantee data plausibility – the highest possible syntactical data quality level.

Generally, this is done even today mostly informally, in rather ad-hoc manners, which is inevitably prone to errors. The process starts with establishing a set of Entity-Relationship diagrams (E-RDs) [1–3], then refining from them a relational db (rdb) model (RDM) scheme [3–5], which is implemented on a version of a Relational Database Management System (RDBMS). Too often, the corresponding software applications that manage these dbs are architectured and developed using UML, user diagrams, classes, agents, ontologies, etc., almost completely decoupled from the business rules.

Generally, missed business rules are discovered only during db software application usage, reported as bugs, and fixed ad-hoc (sometimes even introducing new bugs). This is why, after decades of IT developing, research, and Computer Science University teaching experience, we introduced a middle conceptual level between the E-R and relational ones, our (Elementary) Mathematical Data Model ((E)MDM) [6], as well as a db constraint-driven methodology for designing and developing db software applications [7].

In this framework, for any subuniverse of discourse, we start by algorithmically discovering and establishing the set of all business rules that govern it, along with the corresponding set of entity and relationship sets and their attributes (see Algorithm *A*0 from [3], Section 2.9).

Then, by using the E-R to (E)MDM translation algorithm [8], we refine the initial models, mainly by formalizing business rules as db constraints, i.e., closed Horn clauses of the first-order predicate calculus with equality (FOPC), but also formalizing E-R types as sets, and their attributes as functions. The first part of this paper presents the algorithm that we use in the next step: translating (E)MDM schemes into relational db ones and associated sets of non-relational constraints. Please recall that there are only 6 types of relational constraints (domain/range, not null, default value, unique key, foreign key, and tuple/check), whereas (E)MDM provides 75 [6].

Relational db schemes are then automatically translatable into corresponding ANSI SQL scripts by using Algorithm *A*8 from [3] (see Fig. 4.1, Section 4.1), which may then be similarly translated into rdbs managed by any desired version of a RDBMS, by the corresponding algorithm from the family *AF*8' from [3] (see exercise 4.3 from subsection 4.11.2).

All these algorithms may not only be manually run, but are also implemented in *MatBase*, our intelligent db and knowledge base management system prototype [9], currently developed with MS VBA in two versions – an Access-based one for small to medium dbs and a SQL Server-based one for large dbs. *MatBase* provides graphic user interfaces (GUIs) for both (E)MDM (which includes Datalog¬ [5]), RDM, and E-RDs, as well as algorithms to translate between them, be it directly or by reverse engineering [3, 8, 10].

For enforcing non-relational constraints, *MatBase* first automatically generates a Windows datasheet form for every table in any db it manages. Then, for most of the 75 (E)MDM constraint types, it automatically generates code into the classes of these forms [11–15]. For the remaining constraints, it invites its users to add their code.

The second part of this paper presents examples of enforcing non-relational constraints from the genealogical trees subuniverse, as well as our main guidelines for enforcing such constraints by using event-driven programming with embedded SQL.

## 2. RELATED WORK

Other related approaches to *MatBase* are based on business rules management (BRM) [16, 17] and their corresponding implemented systems (BRMS) and decision managers (e.g., [18–20]). From this perspective, (E)MDM is also a formal BRM, and *MatBase* is an automatically code-generating BRMS.

(E)MDM is also a 5th generation programming language [21, 22] and *MatBase* is also a tool for transparent programming while modeling data at conceptual levels [23].

## 3. *MatBase* PSEUDOCODE ALGORITHM FOR TRANSLATING (E)MDM SCHEMES INTO RELATIONAL ONES AND ASSOCIATED SETS OF NON-RELATIONAL CONSTRAINTS

Figures 1 to 5 show the *MatBase M-R* pseudocode algorithm (referred to as *A*7 in [3]) for translating (E)MDM schemes into relational ones and associated sets of non-relational constraints.

```
Method addColumn(R, e)
add column e to table R, having domain constraint (i.e., data type
      and, if any, range [minValue, maxValue]) corresponding to e's
      codomain;
if e is computed then add corresponding computation trigger;
End Method addColumn;
```

Figure 1: Method *addColumn* of Algorithm *M-R*

```
Method addForeignKey(R, A, S)
add a numeric column A to table R;
if S is a table then
  declare A as foreign key having maximum value corresponding to S's
  cardinality restriction, and referencing primary key x of table S;
else code S' values numerically and add corresponding domain
  constraint to A;
end if;
End Method addForeignKey;
```

Figure 2: Method *addForeignKey* of Algorithm *M-R*

## 4. AN EXAMPLE

Applying the *M-R* algorithm to the (E)MDM scheme from [7, 10, 24], the relational scheme shown in Tables 1 to 7 and the associated sets of non-relational constraints shown in Figures 6 to 9 are obtained.

In fact, because of its unexpected limitation, for MS Access we had to change the type of the *Age* column from *Calculated* to NAT (and to subsequently add code for computing its values), as, although syntactically correct, it does not accept as expression IsNull(PassedAwayYear, Year(Date()), PassedAwayYear) – BirthYear for calculated fields.

To make table schemes fully understandable, we added to the relational schemes examples of plausible instances. We recall our convention for table schemes [3]:

```
                    Method completeScheme(R)
loop for all structural functions f from R to S, except for set
   inclusion-type ones (i.e., canonical injections)
   addForeignKey(R, f, S);
end loop;
loop for all attributes a of R, except for the object identifier x
  addColumn(R, a);
end loop;
loop for all totality constraints c associated to R
   add to R's scheme corresponding NOT NULL constraints;
end loop;
loop for all keys k associated to R
   add to R's scheme corresponding k key constraint;
end loop;
loop for all object constraints t associated to R involving only map-
       pings defined over R and one universally quantified variable
   add to R's scheme corresponding tuple constraint t;
end loop;
End Method completeScheme;
```

Figure 3: Method *completeScheme* of Algorithm *M-R*

- Table title contains its name, followed in parentheses by its (unique) keys (with the primary one underlined), and then by its tuple (check) constraints.
- First header line contains the column names.
- Second header line contains the column ranges (which, for foreign keys, are the images, i.e., the value sets, of the corresponding referenced columns) or, for computed ones, their computation formula.
- Third (and final) header line contains the corresponding NOT NULL constraints.

## 5. ENFORCING THE ASSOCIATED NON-RELATIONAL CONSTRAINT SETS

All non-relational constraints are enforced in the classes of the Windows forms generated over the corresponding relational db tables. For some of them, only SQL code is needed, but most of them need VBA/C# code with embedded SQL. For most types of such constraints, *MatBase* automatically generates code for their enforcement that calls its public methods from the *Constraint* library [9]; for the remaining constraints, especially the object-type ones, db software architects need to develop it. *MatBase*'s SQL Server version also generates T-SQL trigger code for enforcing the non-relational constraints at the db level as well, for protecting data integrity even if the client app is circumvented.

Fortunately, thanks to clever null values logics, both MS Access and SQL Server may, however, enforce some non-relational constraints as well. For example, $C_{17}$: $(\forall x \in MARRIAGES)(MarriageYear(x) \notin \text{NULLS} \Rightarrow DivorceYear(x) \notin \text{NULLS} \vee DivorceYear(x) \geq MarriageYear(x))$ (nobody may divorce somebody before marrying him/her) may be enforced by the tuple (check) constraint *DivorceYear* ≥ *MarriageYear*; similarly, $C_{24}$: $(\forall x \in REIGNS)(ToY(x) \in \text{NULLS} \vee ToY(x) \geq FromY(x))$ (no reign may end before starting) may be enforced by the tuple (check) constraint *ToY* ≥ *FromY*.

```
Method createTable(R)
if there is no table R and R has attributes, or is referencing, or it
  is not referencing, but it is not static and with small cardinality
then create table R;
  if R is a subset of S then
    AddForeignKey(R, x, S);
    declare x as primary key;
    loop for all other sets T with R ⊆ T
      AddForeignKey(R, xT, T);
      add to R's scheme NOT NULL and UNIQUE constraints for xT;
    end loop;
  else add an autonumber primary key x, having maximum value
      corresponding to R's cardinality restriction;
  end if;
  completeScheme(R);
end if;
End Method createTable;
```

Figure 4: Method *createTable* of Algorithm *M-R*

Dually, we need to take into consideration the limitations of the underlying RDBMS, which sometimes force us to enforce relational-type constraints through code as well. For example, unfortunately, MS SQL Server accepts for the column *RULERS.Age* the computation definition AS isnull(PassedYear, year(getdate()))) – BirthYear, but only as a virtual column, not a persisted one, as (due to getdate()) this expression is non-deterministic; consequently, constraint $C_6$: $(\forall x \in RULERS)(Sex(x) \neq \text{'N'} \Rightarrow 0 \leq Age(x) \leq 140)$ (anybody's age must be a natural at most equal to 140) must also be enforced through code, just like for MS Access, although it is of relational domain/range (check) type. Fig. 10 shows the corresponding VBA code. Please note that $C_6$ makes the tuple constraint *BirthYear* ≤ *PassedAwayYear* redundant.

Constraints $C_7$, $C_8$, $C_{18}$, and $C_{19}$ may be very easily enforced only through adding a WHERE clause in four SELECT SQL statements as follows:

- In the *RULERS* and *MARRIAGES* Windows forms, as *RULERS* has the unique key *Name* • *BirthYear* • *Dynasty*, *MatBase* generates for both foreign keys *Mother*, *Father*, *KilledBy*, *Wife*, and *Husband* combo-boxes whose row sources are computed by the following statement:

```
Algorithm M-R (Translation of (E)MDM schemes into relational ones
        and associated sets of non-relational constraints)
-----------------------------------------------------------------------
Input: a(n) (E)MDM db scheme M.
Output: corresponding relational db scheme R and associated set of
   non-relational constraints C.
Strategy:
-----------------------------------------------------------------------
loop for all entity-type object sets E in M (in bottom-up order, from
    non-referencing object sets to non-referenced, only referencing
    ones)
   if E is not a computed set then createTable(E);
   else add a corresponding view E to the db scheme;
end loop;
loop for all relationship-type object sets R in D (in bottom-up order,
    from only referenced, non-referencing object sets to non-
    referenced, only referencing ones)
   createTable(R);
   loop for all R's roles (canonical Cartesian projections) r
      addForeignKey(R, r, S);
      add to R's scheme a NOT NULL constraint for foreign key r;
   end loop;
end loop;
add all unused (non-relational) constraints to the non-relational
   constraint set C;
End Algorithm M-R;
```

Figure 5: Algorithm *M-R* (Translate (E)MDM schemes to RDM ones and associated sets of non- relational constraints)

```
SELECT RULERS.x, [Name] & ", " & [BirthYear] & IIf(IsNull([RULERS].[Dynasty]),"",", " &
        [DYNASTIES].[Dynasty]) AS [Name, BirthYear, Dynasty]
FROM DYNASTIES RIGHT JOIN RULERS ON DYNASTIES.x = RULERS.Dynasty
ORDER BY [Name] & ", " & [BirthYear] & IIf(IsNull([RULERS].[Dynasty]),"",", " &
      [DYNASTIES].[Dynasty]);
```

Table 1: *COUNTRIES*

***COUNTRIES*** (x, *Country*, *Capital*)

| x | *Country* | *Capital* |
|---|---|---|
| auto(3) | ASCII(255) | *Im*(*CITIES.x*) |
| NOT NULL | NOT NULL | |
| 1 | U.K. | 1 |
| 2 | France | 2 |
| 3 | U.S.A. | |

Table 2: *CITIES*

***CITIES*** (x, *City* • *Country*)

| x | *City* | *Country* |
|---|---|---|
| auto(6) | ASCII(255) | *Im*(*COUNTRIES.x*) |
| NOT NULL | NOT NULL | NOT NULL |
| 1 | London | 1 |
| 2 | Paris | 2 |
| 3 | Sandringham | 1 |
| 4 | Reading | 1 |
| 5 | Los Angeles | 3 |

$C_2$: *Country* ° *Capital* null-reflexive (The capital city of any country must belong to that country.)

Figure 6: Non-relational constraint set associated with tables *COUNTRIES* and *CITIES*

Table 3: *DYNASTIES*

***DYNASTIES*** (x, *Dynasty*, *Founder*)

| x | *Dynasty* | *Country* | *Founder* |
|---|---|---|---|
| auto(8) | ASCII(255) | *Im*(*COUNTRIES.x*) | *Im*(*RULERS.x*) |
| NOT NULL | NOT NULL | NOT NULL | |
| 1 | Windsor | 1 | |

Table 4: *TITLES*

***TITLES*** (x, *Title*)

| x | *Title* |
|---|---|
| auto(2) | ASCII(32) |
| NOT NULL | NOT NULL |
| 1 | King |
| 2 | Queen |
| 3 | Prince |
| 4 | Princess |
| 5 | Duchess |

- For $C_7$ and $C_{19}$, injecting to it the clause WHERE Sex = 'F' for *Mother* and *Wife* will select in them only female rulers (see, e.g., Fig. 11).

- For $C_8$ and $C_{18}$, injecting to it the clause WHERE Sex = 'M' for *Father* and *Husband* will select in them only male rulers (see, e.g., Fig. 12).

A partially similar approach may be used for enforcing constraint $C_2$: *Country* ° *Capital* null-reflexive (the capital city of any country must belong to that country), which might be violated either by choosing for a country a city from another one or/and by changing the country to which a capital city belongs.

As *CITIES* has the unique key *City* • *Country*, *MatBase* generates for the row source of the *Capital* combo-box from the *COUNTRIES* Windows form the following SQL statement:

```
SELECT CITIES.x, CITIES.City, COUNTRIES.Country
FROM COUNTRIES INNER JOIN CITIES ON COUNTRIES.x = CITIES.Country
ORDER BY COUNTRIES.Country, CITIES.City;
```

This allows users to choose capitals for any country from the whole set of known cities, as shown on Fig. 13.

To prevent users from violating $C_2$ when choosing capital cities for countries, it is enough to replace this SQL statement with the following dynamic one, where $x$ is the value of the primary key column of *COUNTRIES*, and run it every time when users enter a new line of this table:

"SELECT x, City FROM CITIES WHERE Country = " & x & " ORDER BY City".

The corresponding VBA code is shown in Fig. 14, and its effect is presented in Fig. 15, where, e.g., for U.K. only cities of U.K. are available to choose its capital from.

Table 5: *RULERS*

***RULERS*** *(x, URL, Name • Dynasty • BirthYear, Mother • Name, Father • Name)*

*BirthYear ≤ PassedAwayYear*

| *x* | *Name* | *Sex* | *BirthYear* | *PassedAwayYear* | *Age* | *Mother* | *Father* | *KilledBy* | *Dynasty* | *Title* | *BirthPlace* | *Nationality* | *PassedAwayPlace* | *URL* |
|---|---|---|---|---|---|---|---|---|---|---|---|---|---|---|
| auto(16) | ASCII(255) | {‘M’, ‘F’, ‘N’} | [-6500, *CurrentYear()*] | [-6500, *CurrentYear()*] | *isNull(PassedAwayYear, CurrentYear()) – BirthYear* | *Im(x)* | *Im(x)* | *Im(x)* | *Im(DYNASTIES.x)* | *Im(TITLES.x)* | *Im(CITIES.x)* | *Im(COUNTRIES.x)* | *Im(CITIES.x)* | ASCII(255) |
| NOT NULL | NOT NULL | NOT NULL | | | | | | | | | | | | |
| 1 | Charles III | ‘M’ | 1948 | | 78 | | | | 1 | 1 | 1 | 1 | | en.wikipedia.org/wiki/Charles_III |
| 2 | Diana Spencer | ‘F’ | 1961 | 1997 | 36 | | | | 1 | 4 | 3 | 1 | 2 | en.wikipedia.org/wiki/Diana,_Princess_of_Wales |
| 3 | William | ‘M’ | 1982 | | 44 | 2 | 1 | | 1 | 3 | 1 | 1 | | en.wikipedia.org/wiki/William,_Prince_of_Wales |
| 4 | Harry | ‘M’ | 1984 | | 42 | 2 | 1 | | 1 | 3 | 1 | 1 | | en.wikipedia.org/wiki/Prince_Harry,_Duke_of_Sussex |
| 5 | Catherine | ‘F’ | 1982 | | 44 | | | | | 4 | 4 | 1 | | en.wikipedia.org/wiki/Catherine,_Princess_of_Wales |
| 6 | Meghan | ‘F’ | 1981 | | 45 | | | | | 5 | 5 | 3 | | en.wikipedia.org/wiki/Meghan,_Duchess_of_Sussex |
| 7 | Camilla | ‘F’ | 1947 | | 79 | | | | 1 | 2 | 1 | 1 | | en.wikipedia.org/wiki/Queen_Camilla |
| 8 | Andrew Parker Bowles | ‘M’ | 1939 | | 87 | | | | | | | 1 | | en.wikipedia.org/wiki/Andrew_Parker_Bowles |

For the second part of $C_2$, users must not be allowed to change the countries they belong to for capital cities. The corresponding VBA code is shown on Fig. 16. An example of the error message displayed by this method is shown in Fig. 17.

Please note that, when inserting data for a new city (i.e., *NewRecord* = *True*), $C_2$ needs not to be checked, as such cities may not be country capitals yet. Moreover, and much more important, $C_2$ implies constraint *Capital* key, as no city might be selected for two countries in the corresponding combo-box: consequently, this unique key may be dropped from the relational db schema, making it more efficient.

$C_{27}$: $Mother : RULERS \rightarrow RULERS$ acyclic (Nobody may be his/her maternal ancestor or descendant.)

$C_{28}$: $Father : RULERS \rightarrow RULERS$ acyclic (Nobody may be his/her paternal ancestor or descendant.)

$C_{36}$: $Mother \bullet Father : RULERS \rightarrow RULERS^2$ acyclic (Nobody may be his/her ancestor or descendant.)

$C_{29}$: $\neg PassedAwayYear \vdash \neg KilledBy$ ("There's no living with a killing.")

$C_4$: $Dynasty \circ Founder$ null-reflexive (The founder of any dynasty must belong to that dynasty.)

$C_5$: $(\forall x \in RULERS)(Dynasty(x) \notin \text{NULLS} \wedge Founder(Dynasty(x)) \notin \text{NULLS} \wedge BirthYear(Founder(Dynasty(x))) \notin \text{NULLS} \Rightarrow PassedAwayYear(x) \in \text{NULLS} \vee PassedAwayYear(x) > BirthYear(Founder(Dynasty(x))))$ (Nobody may belong to a dynasty that was founded after his/her death).

$C_6$: $(\forall x \in RULERS)(Sex(x) \neq \text{'N'} \Rightarrow 0 \leq Age(x) \leq 140)$ (Age must be a natural at most equal to 140.)

$C_7$: $(\forall x \in RULERS)(Sex(Mother(x)) = \text{'F'})$ (Mothers' sex must be 'F'.)

$C_8$: $(\forall x \in RULERS)(Sex(Father(x)) = \text{'M'})$ (Fathers' sex must be 'M'.)

$C_9$: $(\forall x \in RULERS)(Sex(x) = \text{'N'} \Rightarrow Mother(x) \in \text{NULLS} \wedge Father(x) \in \text{NULLS} \wedge Dynasty(x) \in \text{NULLS})$ (Non-persons may not have parents or belong to dynasties.)

$C_{12}$: $(\forall x \in RULERS)(BirthYear(x) \notin \text{NULLS} \wedge Mother(x) \notin \text{NULLS} \wedge BirthYear(Mother(x)) \notin \text{NULLS} \wedge Sex(x) \neq \text{'N'} \Rightarrow 5 \leq BirthYear(x) - BirthYear(Mother(x)) \leq 75 \wedge (PassedAwayYear(Mother(x)) \notin \text{NULLS} \Rightarrow (BirthYear(x) \leq PassedAwayYear(Mother(x)))$ (Women may give birth only between 5 and 75 years, but before passing away.)

$C_{13}$: $(\forall x \in RULERS)(BirthYear(x) \notin \text{NULLS} \wedge Father(x) \notin \text{NULLS} \wedge BirthYear(Father(x)) \notin \text{NULLS} \wedge Sex(x) \neq \text{'N'} \Rightarrow 9 \leq BirthYear(x) - BirthYear(Father(x)) \leq 100 \wedge (PassedAwayYear(Father(x)) \notin \text{NULLS} \Rightarrow (BirthYear(x) \leq PassedAwayYear(Father(x)) + 1))$ (Men may have children only between 9 and 100 years, and at most one year after passing away.)

$C_{14}$: $(\forall x \in RULERS)(KilledBy(x) \notin \text{NULLS} \Rightarrow (BirthYear(KilledBy(x)) \notin \text{NULLS} \Rightarrow BirthYear(KilledBy(x)) \leq PassedAwayYear(x)) \wedge PassedAwayYear(x) \leq isNull(PassedAwayYear(KilledBy(x)), CurrentYear()))$ (Any killer must have been alive when his/her victim was killed.)

Figure 7: Non-relational constraint set associated with table *RULERS*

Constraint $C_4$: *Dynasty* ° *Founder* null-reflexive (the founder of any dynasty must belong to that dynasty) is of the same type as $C_2$, so it may be enforced similarly, with a *Form_Current* method in the *DYNASTIES* class and a *Dynasty_BeforeUpdate* one in the *RULERS* class; the only difference should be that, in the *Form_Current* method, the dynamic SQL query must select not only the persons belonging to the current dynasty, but also those not belonging to any dynasty. Fortunately, as, generally, $f \circ g$ (null-)reflexive implies $g$ one-to-one, $C_4$ makes "*Founder* key" redundant.

Constraint $C_9$: $(\forall x \in RULERS)(Sex(x) = \text{'N'} \Rightarrow Mother(x) \in \text{NULLS} \wedge Father(x) \in \text{NULLS} \wedge Dynasty(x) \in \text{NULLS})$ (non-persons may not have parents or belong to dynasties) may be elegantly enforced by the code shown in Fig. 18: method *Form_Current* blocks updating these 4 columns for lines having *Sex* = 'N' and allows it for 'M' and 'F', while the *Sex_AfterUpdate* one nullifies their values whenever *Sex* is changed to 'N', also displaying the warning message shown in Fig. 19.

Table 6: *MARRIAGES*

***MARRIAGES*** (x, *Husband* • *Wife* • *MarriageYear*, *Husband* • *Wife* • *DivorceYear*)

$MarriageYear \leq DivorceYear$

| x | *MarriageYear* | *DivorceYear* | *Husband* | *Wife* |
|---|---|---|---|---|
| auto(16) | [-6500, *CurrentYear*()] | [-6500, *CurrentYear*()] | *Im*(*RULERS.x*) | *Im*(*RULERS.x*) |
| NOT NULL | | | NOT NULL | NOT NULL |
| 1 | 1981 | 1996 | 1 | 2 |
| 2 | 2005 | | 1 | 7 |
| 3 | 2011 | | 3 | 5 |
| 4 | 2018 | | 4 | 6 |
| 5 | 1973 | 1995 | 8 | 7 |

Object constraints $C_{12}$ to $C_{15}$, as well as $C_{20}$, $C_{21}$, $C_{25}$, $C_{26}$, and $C_{29}$ to $C_{33}$ may also be enforced through VBA *Form_BeforeUpdate* and *column_BeforeUpdate* (.Net *Validating*) methods, like $C_6$ (see Fig. 10) and $C_2$ (see Fig. 16). $C_{37}$ may be enforced through the VBA *PassedAwayYear_AfterUpdate* (.Net *Validated*) method (its □ symbol is the "***always** in every future state*" quantifier from the temporal first-order logic).

*MatBase* automatically generates corresponding code in the *RULERS* class for enforcing both the acyclicity constraints $C_{27}$ and $C_{28}$ (see [13]) and the anti-existence $C_{29}$ one.

## 6. DISCUSSION

*Proposition* (Characterization of Algorithm *M–R*) Algorithm *M-R* (Figures 1 to 5) is:

(i) linear (i.e., has complexity $O(n)$, $n$ natural).
(ii) sound
(iii) complete
(iv) optimal.

*Proof*:

(i) (*complexity*) Let $e$ be the number of entity-type sets of scheme M, $r$ the one of relationship-type ones, $a$ the number of all their attributes, $f = ccp + nccp$ the one of their structural functions (with $ccp$ canonical Cartesian projections and $nccp$ non-canonical Cartesian projections), and $c$

= *rc* + *nrc* the one of associated constraints (with *rc* relational and *nrc* non-relational); the first loop of *M-R* is executed *e* times, calling method *createTable* at most *e* times in total; the second loop is executed *r* times, calling method *createTable r* times and generating *ccp* foreign keys and corresponding not null constraints; finally, its third loop is executing *nrc* times; consequently, method *createTable* is called at most *e* + *r* times and, in total, creates the tables of the output relational scheme *R* (at most *e* + *r*), their primary key constraints, their *nccp* foreign key columns not corresponding to canonical Cartesian projections (including those corresponding to canonical inclusions), as well as calling in the end the method *completeScheme*; this method adds in its first loop the *nccp* foreign key constraints, through calls to method *addColumn* all *a* attributes, all their domain (range), not null, unique key, and tuple (check) constraints, i.e., all rc relational constraints of *R*; as methods *addColumn* and *addForeignKey* do not have loops, it follows that, in total, *M-R* performs exactly *n* = *e* + *r* + *f* + *c* steps, hence it never loops infinitely and is linear in the sum of the total numbers of sets, functions, and constraints of the input (E)MDM scheme *M*.

(ii) (*soundness*) As *M-R* outputs only relational tables, views, constraints, and sets of associated non-relational constraints, it is sound.

(iii) (*completeness*) As *M-R* accepts as inputs any (E)MDM scheme, it is also complete.

(iv) (*optimality*) As *M-R* treats any input set, function, and constraint only once and generates any table, view, column, and constraint only once, it is also optimal. *Q.E.D.*

For the above example, of course, *e* = 7 (as there are 7 entity-type sets, *COUNTRIES, CITIES, DYNASTIES, TITLES, RULERS, MARRIAGES*, and *REIGNS*, all of them translated into corresponding rdb tables), *r* = *ccp* = 0 (there is no relationship-type set), *a* = 21 (with the 7 object identifiers translated into surrogate primary keys *x*, *Country* from *COUNTRIES*, *City* from *CITIES*, *Dynasty* from *DYNASTIES*, *Title* from *TITLES*, *Name, Sex, BirthYear, PassedAwayYear, Age*, and *URL* from *RULERS, MarriageYear* and *DivorceYear* from *MARRIAGES*, *FromY* and *ToY* from *REIGNS*), *f* = *nccp* = 17 (*Capital* : *COUNTRIES* ↔ *CITIES*, *City* : *CITIES* → *COUNTRIES*, *Country* : *DYNASTIES* → *COUNTRIES*, *Founder* : *DYNASTIES* ↔ *RULERS*, *Mother* : *RULERS* → *RULERS*, *Father* : *RULERS* → *RULERS*, *KilledBy* : *RULERS* → *RULERS*, *Dynasty* : *RULERS* → *DYNASTIES*, *Title* : *RULERS* → *TITLES*, *BirthPlace* : *RULERS* → *CITIES*, *PassedAwayPlace* : *RULERS* → *CITIES*, *Nationality* : *RULERS* → *COUNTRIES*, *Husband* : *MARRIAGES* → *RULERS*, *Wife* : *MARRIAGES* → *RULERS*, *Ruler* : *REIGNS* → *RULERS*, *Country* : *REIGNS* → *COUNTRIES*, and *Title* : *REIGNS* → *TITLES*, all of them translated into corresponding foreign keys), *rc* = 21 + 13 + 14 + 17 + 14 + 3 = 82 (with 7 primary keys and their 7 domain and 7 not null associated constraints, other 13 not null ones, one for *COUNTRIES* and *TITLES*, 2 for *CITIES*, *DYNASTIES*, *RULERS*, and *MARRIAGES*, and 3 for *REIGNS*, 14 domain ones for the rest of the attributes, 17 foreign keys, the 14 unique keys *Country* and *Capital* of *COUNTRIES*, *City* • *Country* of *CITIES, Dynasty* and *Founder* of *DYNASTIES*, *Title* of *TITLES*, *URL*, *Name* • *Dynasty* •*BirthYear*, *Name* • *Mother*, and *Name* • *Father* of *RULERS*, *Husband* • *Wife* • *MarriageYear* and *Husband* • *Wife* • *DivorceYear* of *MARRIAGES*, *Ruler* • *Country* • *FromY* and *Ruler* • *Country* • *ToY* from *REIGNS*, as well as the 3 tuple ones, *BirthYear* ≤ *PassedAwayYear*, *MarriageYear* ≤ *DivorceYear* and *FromY* ≤ *ToY*), and *nrc* = 27 (the one from Figure 6, the 13 from Figure 7, 8 from Figure 8, and 5 from Figure 9). In total, algorithm *M-R* performs *n* = 7 + 21 + 17 + 82 + 27 = 154 steps to translate this (E)MDM scheme.

Fortunately, as $C_2$ implies the key *Capital*, $C_4$ implies the key *Founder*, $C_{33}$ implies the tuple constraint *MarriageYear* ≤ *DivorceYear*, and $C_{34}$ implies the keys *Ruler* • *Country* • *FromY* and *Ruler* • *Country* • *ToY*, there are only 77 relational constraints to be enforced. Unfortunately, although $C_6$ implies the tuple constraint *BirthYear* ≤ *PassedAwayYear* for physical persons (i.e., having *Sex* 'F' or 'M'), but MS check constraints do not accept either $Sex = \text{'N'} \Rightarrow BirthYear \leq PassedAwayYear$ or *if* $Sex = \text{'N'}$ then $BirthYear \leq PassedAwayYear$, this tuple constraint must be enforced in the rdb scheme.

$C_{18}$: $(\forall x \in MARRIAGES)(Sex(Wife(x)) = \text{'F'})$ (Wives' sex must be 'F'.)

$C_{19}$: $(\forall x \in MARRIAGES)(Sex(Husband(x)) = \text{'M'})$ (Husbands' sex must be 'M'.)

$C_{20}$: $(\forall x \in MARRIAGES)(MarriageYear(x) \notin \text{NULLS} \Rightarrow ((BirthYear(Husband(x)) \in \text{NULLS} \vee BirthYear(Husband(x)) \leq MarriageYear(x)) \wedge (PassedAwayYear(Husband(x)) \in \text{NULLS} \vee PassedAwayYear(Husband(x)) \geq MarriageYear(x))) \wedge (BirthYear(Wife(x)) \in \text{NULLS} \vee BirthYear(Wife(x)) \leq MarriageYear(x)) \wedge (PassedAwayYear(Wife(x)) \in \text{NULLS} \vee PassedAwayYear(Wife(x)) \geq MarriageYear(x))))$ (Nobody may get married before being born or after death.)

$C_{21}$: $(\forall x \in MARRIAGES)(DivorceYear(x) \notin \text{NULLS} \Rightarrow ((BirthYear(Husband(x)) \in \text{NULLS} \vee BirthYear(Husband(x)) \leq DivorceYear(x)) \wedge (PassedAwayYear(Husband(x)) \in \text{NULLS} \vee PassedAwayYear(Husband(x)) \geq DivorceYear(x))) \wedge (BirthYear(Wife(x)) \in \text{NULLS} \vee BirthYear(Wife(x)) \leq DivorceYear(x)) \wedge (PassedAwayYear(Wife(x)) \in \text{NULLS} \vee PassedAwayYear(Wife(x)) \geq DivorceYear(x))))$ (Nobody may divorce before being born or after death.)

$C_{30}$: $(\forall x \in MARRIAGES)(BirthYear(Husband(x)) \leq BirthYear(Wife(x)) \leq isNull(PassedAwayYear(Husband(x)), CurrentYear()) \vee BirthYear(Wife(x)) \leq BirthYear(Husband(x)) \leq isNull(PassedAwayYear(Wife(x)), CurrentYear()) \vee BirthYear(Husband(x)) \leq PassedAwayYear(Wife(x)) \leq isNull(PassedAwayYear(Husband(x)), CurrentYear()) \vee BirthYear(Wife(x)) \leq PassedAwayYear(Husband(x)) \leq isNull(PassedAwayYear(Wife(x)), CurrentYear()))$ (For any marriage, husband and wife must have been simultaneously alive at least one day.)

$C_{31}$: $(\forall x,y \in MARRIAGES)(x \neq y \wedge (MarriageYear(y) < MarriageYear(x) < isNull(DivorceYear(y), CurrentYear()) \vee MarriageYear(x) < MarriageYear(y) < isNull(DivorceYear(y), CurrentYear() \vee MarriageYear(y) < DivorceYear(x) < isNull(DivorceYear(y), CurrentYear()) \vee MarriageYear(x) < DivorceYear(y) < isNull(DivorceYear(x), CurrentYear()))) \Rightarrow Husband(x) \neq Husband(y) \wedge Wife(x) \neq Wife(y))$ (No spouse may remarry while being married.)

$C_{32}$: $(\forall x \in MARRIAGES)(Father(Husband(x)) \neq Father(Wife(x)) \wedge Mother(Husband(x)) \neq Mother(Wife(x)) \wedge Father(Wife(x)) \neq Husband(x) \wedge Mother(Husband(x)) \neq Wife(x))$ (Incestuous marriages are forbidden.)

$C_{33}$: $(\forall x \in MARRIAGES)(MarriageYear(x) \notin \text{NULLS} \wedge PassedAwayYear(Husband(x)) \in \text{NULLS} \wedge PassedAwayYear(Wife(x)) \in \text{NULLS} \Rightarrow 0 \leq isNull(DivorceYear(y), CurrentYear()) - MarriageYear(x) \leq 140)$ (Marriage years must be naturals at most equal to 140.)

Figure 8: Non-relational constraint set associated with table *MARRIAGES*

Moreover, in [24] we have shown that both *Mother* • *Father* and *Husband* • *Wife* are irreflexive, asymmetric, transitive, intransitive, inEuclidean, and acyclic, but, except for $C_{36}$, all these other 11 non-relational constraints are implied by $C_7 \wedge C_8$ and $C_{18} \wedge C_{19}$, respectively, so that we did not consider them here.

Table 7: *REIGNS*

***REIGNS*** ($\underline{x}$, *Ruler • Country • FromY*, *Ruler • Country • ToY*) *ToY* ≥ *FromY*

| $\underline{x}$ | *FromY* | *ToY* | *Ruler* | *Country* | *Title* |
|---|---|---|---|---|---|
| auto(16) | [-6500, *CurrentYear*()] | [-6500, *CurrentYear*()] | *Im*(*RULERS.x*) | *Im*(*COUNTRIES.x*) | *Im*(*TITLES.x*) |
| NOT NULL | NOT NULL | | NOT NULL | NOT NULL | |
| 1 | 2022 | | 1 | 1 | 1 |
| 2 | 2022 | | 7 | 1 | 2 |

$C_{25}$: $(\forall x \in \mathit{REIGNS})((\mathit{BirthYear}(\mathit{Ruler}(x)) \notin \text{NULLS} \Rightarrow \mathit{BirthYear}(\mathit{Ruler}(x)) \leq \mathit{FromY}(x)) \wedge (\mathit{PassedAwayYear}(\mathit{Ruler}(x)) \notin \text{NULLS} \Rightarrow \mathit{ToY}(x) \notin \text{NULLS} \wedge \mathit{PassedAwayYear}(\mathit{Ruler}(x)) \geq \mathit{ToY}(x)))$ (Nobody may reign before being born or after death.)

$C_{26}$: $(\forall x, y \in \mathit{REIGNS})(x \neq y \wedge \mathit{Country}(x) = \mathit{Country}(y) \wedge \mathit{Ruler}(x) \neq \mathit{Ruler}(y) \wedge (\mathit{FromY}(x) \leq \mathit{FromY}(y) < \mathit{isNull}(\mathit{ToY}(x), \mathit{CurrentYear}()) \vee \mathit{FromY}(y) \leq \mathit{FromY}(x) < \mathit{isNull}(\mathit{ToY}(y), \mathit{CurrentYear}()) \vee \mathit{FromY}(x) < \mathit{ToY}(y) \leq \mathit{isNull}(\mathit{ToY}(x), \mathit{CurrentYear}()) \vee \mathit{FromY}(y) < \mathit{ToY}(x) \leq \mathit{isNull}(\mathit{ToY}(y), \mathit{CurrentYear}())) \Rightarrow \mathit{Sex}(x) = \text{'N'} \vee \mathit{Sex}(y) = \text{'N'} \vee (\mathit{Father}(\mathit{Ruler}(y)) = \mathit{Ruler}(x) \vee \mathit{Father}(\mathit{Ruler}(x)) = \mathit{Ruler}(y) \vee \mathit{Mother}(\mathit{Ruler}(y)) = \mathit{Ruler}(x) \vee \mathit{Mother}(\mathit{Ruler}(x)) = \mathit{Ruler}(y)) \vee (\exists z \in \mathit{MARRIAGES})(\mathit{Husband}(z) = \mathit{Ruler}(x) \wedge \mathit{Wife}(z) = \mathit{Ruler}(y) \vee \mathit{Husband}(z) = \mathit{Ruler}(y) \wedge \mathit{Wife}(z) = \mathit{Ruler}(x)))$ (No country may be simultaneously ruled by 2 persons, except for cases where at least one of them has sex 'N', or the two were married, or parent and child.)

$C_{34}$: $(\forall x, y \in \mathit{REIGNS})(x \neq y \wedge \mathit{Country}(x) = \mathit{Country}(y) \wedge (\mathit{FromY}(x) <= \mathit{FromY}(y) < \mathit{isNull}(\mathit{ToY}(x), \mathit{CurrentYear}()) \vee \mathit{FromY}(y) \leq \mathit{FromY}(x) < \mathit{isNull}(\mathit{ToY}(y), \mathit{CurrentYear}() \vee \mathit{FromY}(x) < \mathit{ToY}(y) \leq \mathit{isNull}(\mathit{ToY}(x), \mathit{CurrentYear}()) \vee \mathit{FromY}(y) < \mathit{ToY}(x) \leq \mathit{isNull}(\mathit{ToY}(y), \mathit{CurrentYear}()) \Rightarrow \mathit{Ruler}(x) \neq \mathit{Ruler}(y))$ (Nobody may reign twice in the same country in the same period.)

$C_{35}$: $(\forall x \in \mathit{REIGNS})(\mathit{Sex}(\mathit{Ruler}(x)) \neq \text{'N'} \Rightarrow \mathit{isNull}(\mathit{ToY}(y), \mathit{CurrentYear}()) - \mathit{FromY}(x) \leq 140)$ (Reign years for males and females must be at most equal to 140.)

$C_{37}$: $(\forall x \in \mathit{PERSONS})(\forall y \in \mathit{REIGNS})(\mathit{PassedAwayYear}(x) \notin \text{NULLS} \wedge \mathit{Ruler}(y) = x \wedge \mathit{ToY}(y) \in \text{NULLS} \Rightarrow \Box(\mathit{ToY}(y) = \mathit{PassedAwayYear}(x)))$ (Any not ended reign must end in the year in which the corresponding ruler passed away.)

Figure 9: Non-relational constraint set associated with table *REIGNS*

The relations between constraints $C_{27}$ (nobody may be his/her maternal ancestor or descendant), $C_{28}$ (nobody may be his/her paternal ancestor or descendant), and $C_{36}$ (nobody may be his/her ancestor or descendant) are the most interesting of them all: obviously, graphs of both *Mother* and *Father* must be trees, i.e., no cycles allowed (which is formalized by $C_{27}$ and $C_{28}$, respectively); however, they do not imply $C_{36}$, as, generally, the graph of such a function is not a tree: e.g., let $m$ = *Parents*($s$, $u$), $s$ = *Father*($v$, $m$), with $\{m, s, u, v\} \subseteq \mathit{PERSONS}$; obviously, although the self-maps *Mother* and *Father* are acyclic, the product *Mother • Father* from $C_{36}$ is not, as $m$ is both the mother and the daughter of $s$. Fortunately, the converse is true, i.e., $C_{36}$ makes both $C_{27}$ and $C_{28}$ redundant.

```
'C6: Anybody's age must be a natural at most equal to 140.
C6:
If Sex = "N" Then GoTo C12  'C6 does not apply for Sex = "N"
If Not IsNull(BirthYear) Then
  Dim RAge, pay As Integer
  pay = IIf(IsNull(PassedAwayYear), Year(Date), PassedAwayYear)
  If pay >= 0 Then
    RAge = IIf(IsNull(PassedAwayYear), Year(Date), PassedAwayYear) _
        - BirthYear
  Else
    RAge = -IIf(IsNull(PassedAwayYear), Year(Date), PassedAwayYear) _
        - BirthYear
  End If
  If RAge < 0 Or RAge > maxLifeYears Then
    Cancel = True
    MsgBox "Please change BirthYear or/and PassedAwayYear, " & _
        "such that age is between 0 and " & maxLifeYears & _
        ": people many not live more than " & maxLifeYears & " years!", _
        vbCritical, "Age for " & Me![Name] & " is " & RAge & "!"
    GoTo exit_point
    End If
End If
'C12: Women may give birth only between 5 and 75 years,
'    and before passing away.
C12:
```

Figure 10: VBA code for enforcing constraint $C_6$

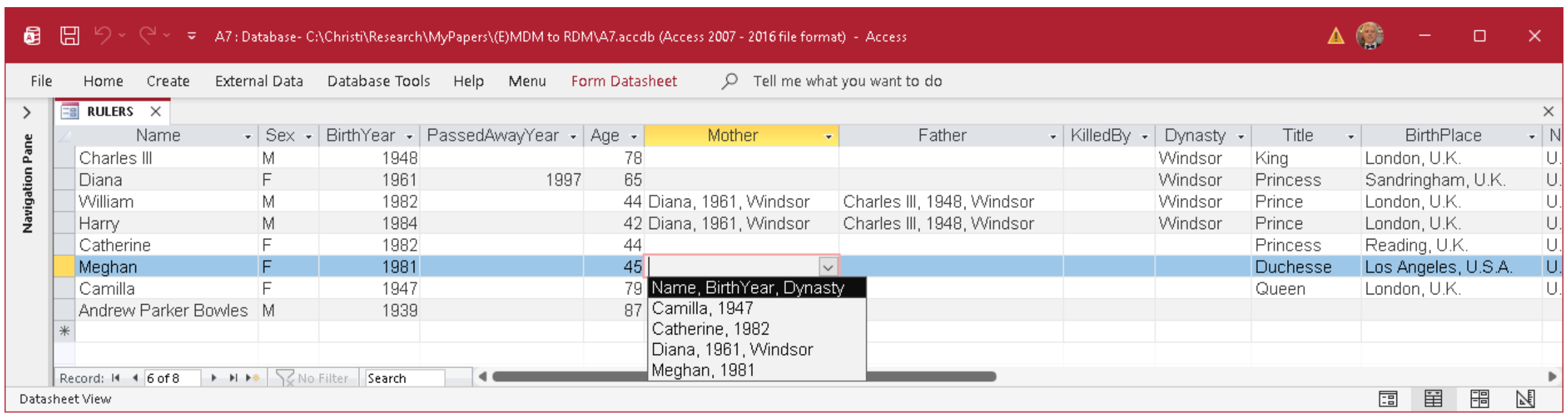

Figure 11: Selecting only females in the *Mother* combo-box of form *RULERS*

Of course, the *M-R* algorithm presented here is used by *MatBase* when importing (E)MDM schemes (from text files). Besides it, *MatBase* also has a corresponding incremental one, embedded in its (E)MDM GUI classes: immediately as users add a new db, it creates the corresponding rdb one; immediately after they add a new set to a db, it creates its corresponding table/view and standard Windows form; as soon as they add a new function defined on a set, it adds the corresponding column to the corresponding table/view and correspondingly updates its Windows form; and, finally, immediately after they add a new constraint to a db, *MatBase* first checks whether it is both syntactically (i.e., from the coherence point of view) and semantically (i.e., from the current db instance point of view) valid [25]; if it is not, *MatBase* rejects it (with a context-dependent error message); otherwise, not only it accepts it, but for relational ones it correspondingly alters the underlying rdb scheme, while for the non-relational ones it either automatically generates code in the corresponding Windows form classes to enforce them or, for the general object-type ones, it invites users to do it.

Please note that, as the *A*9 and *AF*9’ algorithms from [3] are also linear, sound, complete, and optimal, the whole process of translating (E)MDM schemes to actual rdbs using *MatBase* is very fast and secure.

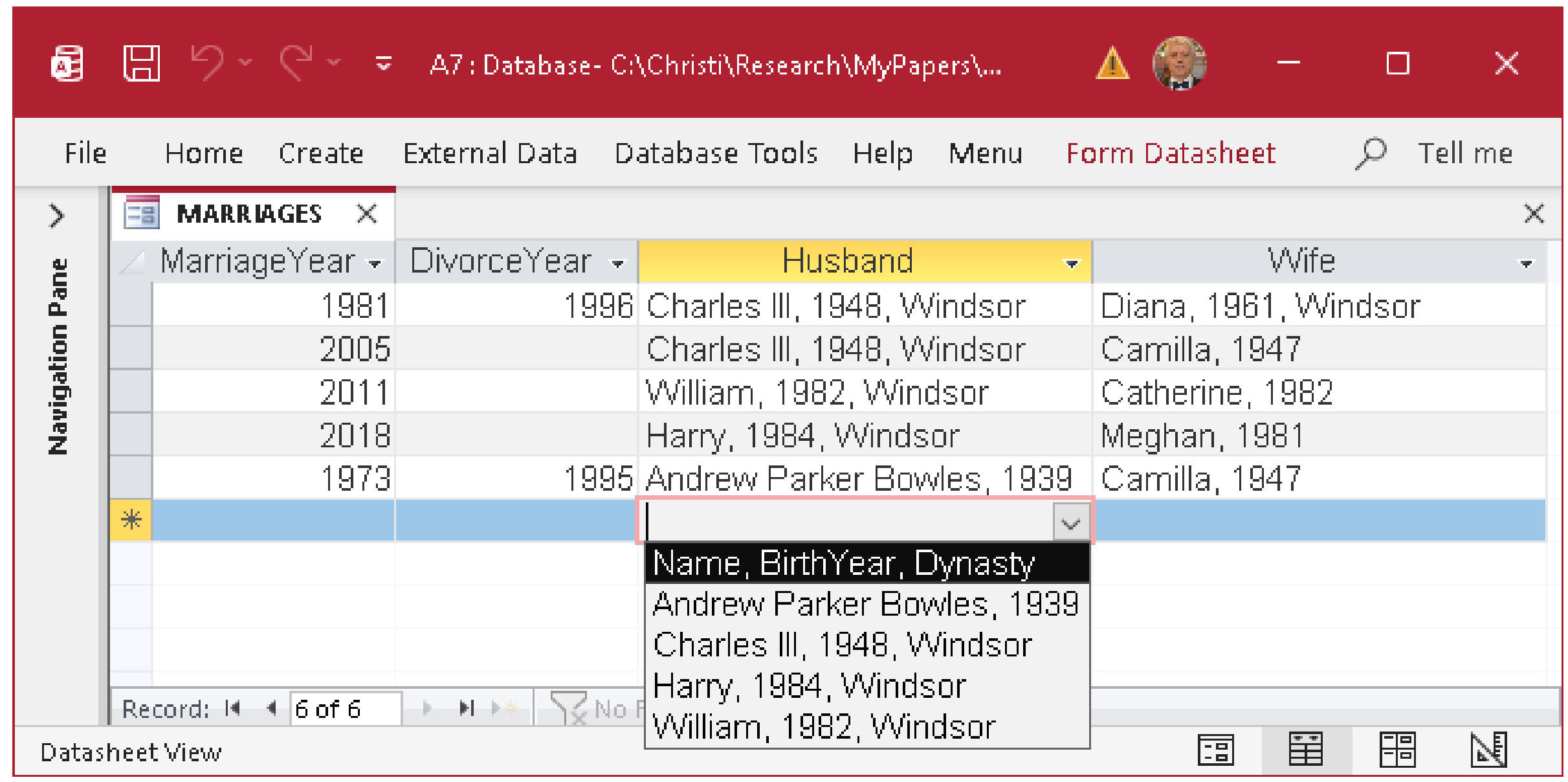

Figure 12: Selecting only males in the *Husband* combo-box of form *MARRIAGES*

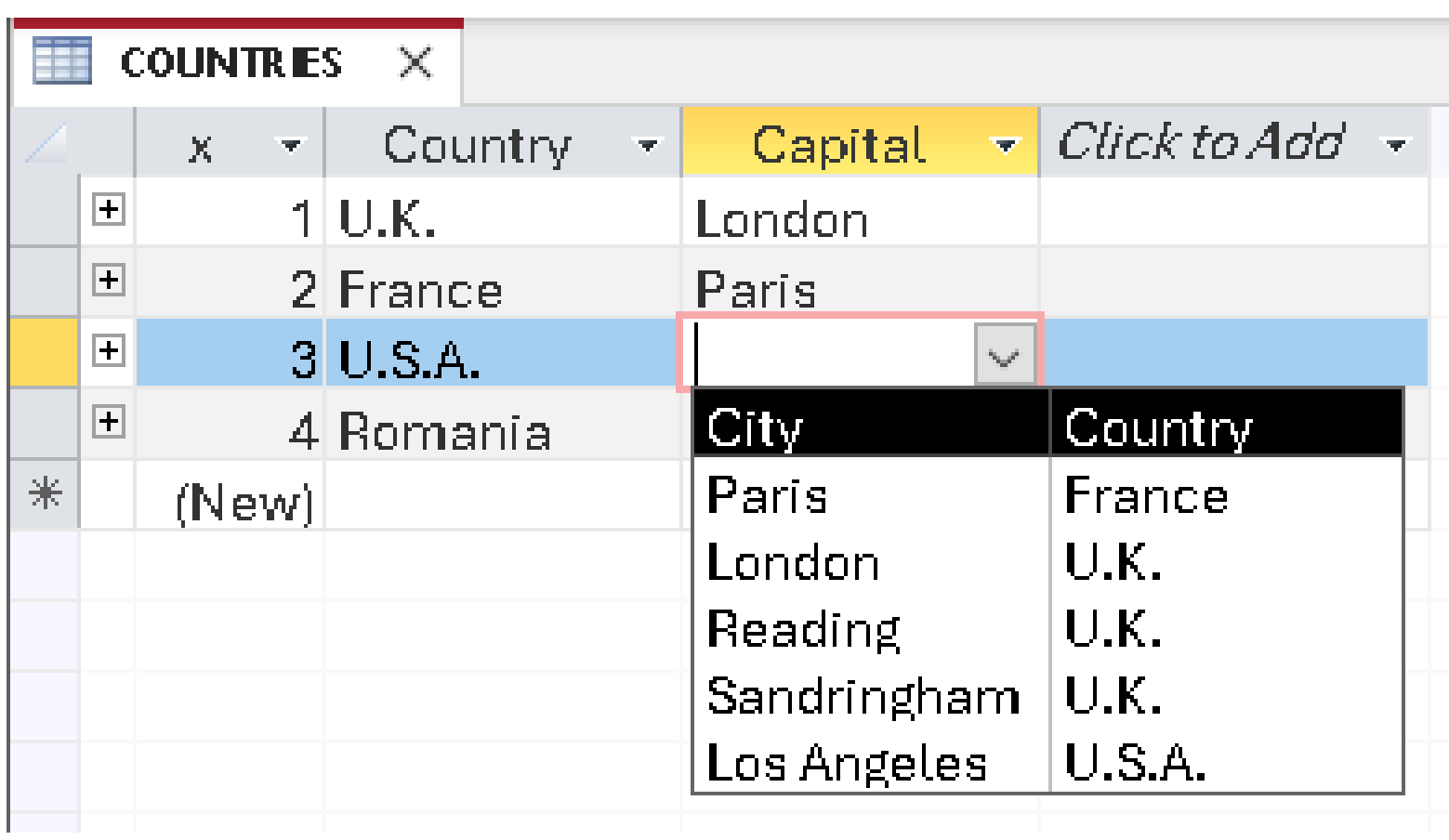

Figure 13: Selecting cities in the *Capital* combo-box of form *COUNTRIES*

```
'COUNTRIES class
Option Compare Database
Option Explicit
_____________________________________________

Private Sub Form_Current()
Capital.RowSource = "SELECT x, City " _
    & "FROM CITIES WHERE Country = " & x _
    & " ORDER BY City"
Capital.Requery
End Sub
```

Figure 14: VBA code for enforcing the first half of constraint $C_2$

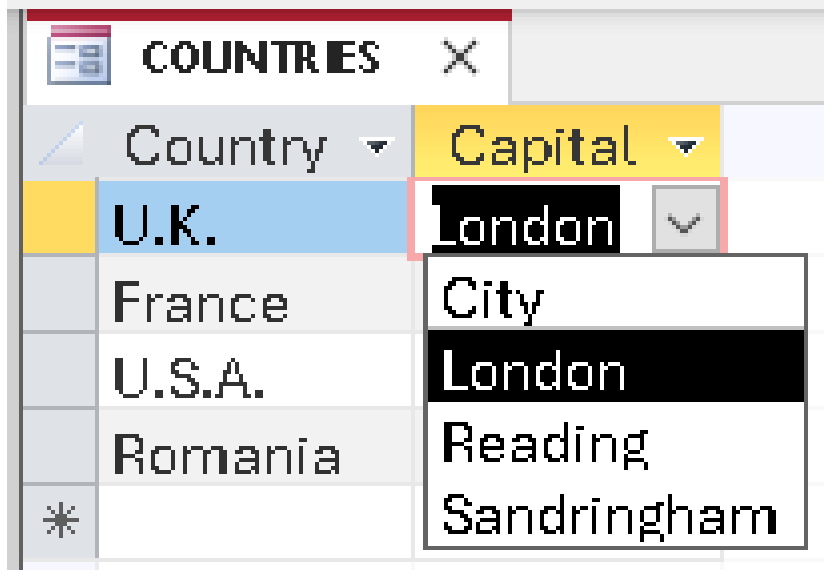

Figure 15: Selecting cities in the *Capital* combo-box of form *COUNTRIES* after enforcing $C_2$

```
'CITIES class
Option Compare Database
Option Explicit

Private Sub Country_BeforeUpdate(Cancel As Integer)
'C2: The capital city of any country must belong to that country.
If Not NewRecord Then
  Dim cntry As Variant
  cntry = DLookup("Country", "COUNTRIES", "Capital = " & x)
  If Not IsNull(cntry) And (IsNull(Country) Or Country <> Country.OldValue) Then
    Cancel = True
    MsgBox "Request denied: you cannot change Country for this city...", _
      vbCritical, City & " is the capital of " & cntry & "!"
    Country.Undo
  End If
End If
End Sub
```

Figure 16: VBA code for enforcing the second half of constraint $C_2$

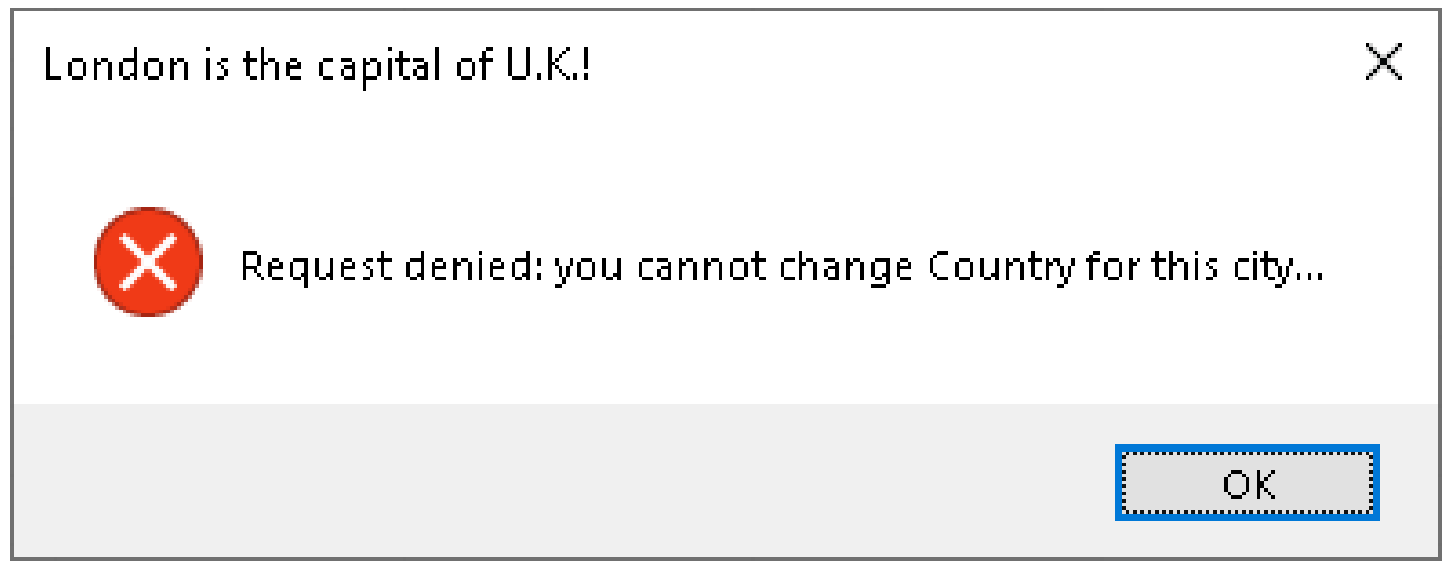

Figure 17: Error message displayed when users attempt to violate constraint $C_2$

## 7. GUIDELINES FOR ENFORCING NON-RELATIONAL CONSTRAINTS

For developing elegant and optimal code to enforce general object-type non-relational constraints we would strongly recommend obeying the following 17 guidelines:

a) Always use our db constraint-driven approach for designing and developing db software applications [7].
b) Always use client software application's event-driven programming languages embedding SQL for enforcing constraints: it does not make sense to pass invalid data to SQL engines when you can reject it from the client side. Also use extended SQL (i.e., T-SQL, PL/SQL, SQL PL, etc.) triggers to guarantee data integrity even when it is updated directly in the db.

For example, Figure 20 shows the T-SQL code generated by *MatBase* in the trigger associated to table *RULERS* for enforcing constraint $C_6$ (also see Figure 10).

```
Private Sub Form_Current()
endReigns = False
'C9: Non-persons may not have parents or belong to dynasties.
If Sex = "N" Then
  Mother.Locked = True
  Father.Locked = True
  Dynasty.Locked = True
Else
  Mother.Locked = False
  Father.Locked = False
  Dynasty.Locked = False
End If
End Sub
```

```
Private Sub Sex_AfterUpdate()
'C9: Non-persons may not have parents or belong to dynasties.
If Sex = "N" Then
  Mother = Null
  Father = Null
  Dynasty = Null
  MsgBox "As not physical persons may not have parents" _
    & " or belong to dynasties, any such not null data" _
    & " for this person has been automatically deleted!", _
    vbExclamation, "Warning!"
  Mother.Locked = True
  Father.Locked = True
  Dynasty.Locked = True
Else
  Mother.Locked = False
  Father.Locked = False
  Dynasty.Locked = False
End If
End Sub
```

Figure 18: VBA code for enforcing non-relational constraint $C_9$

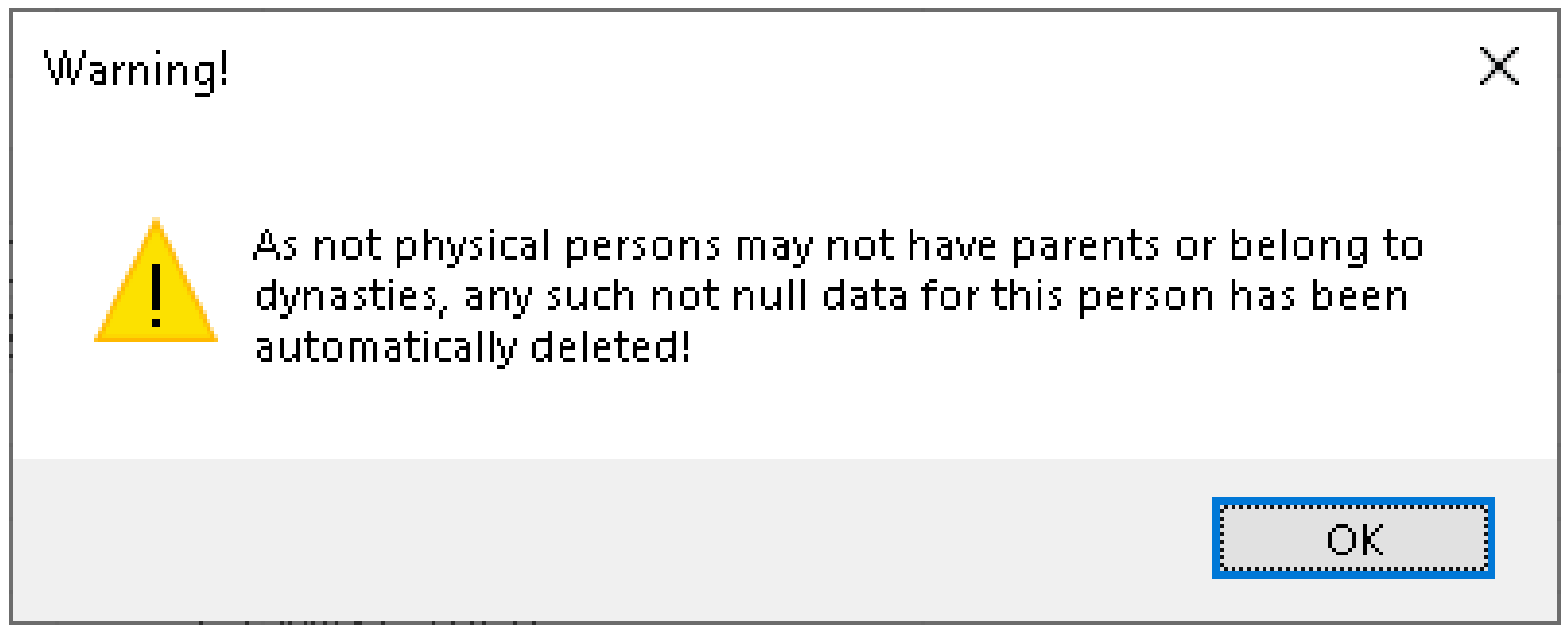

Figure 19: Warning message displayed when users attempt to violate constraint $C_9$

c) You must use at least one method in any Windows form class for which the corresponding constraint has a quantifier or a function defined on a corresponding set. For example, as $C_{26}$ contains two quantifiers, namely ($\forall x,y \in REIGNS$) and ($\exists z \in MARRIAGES$), you need for it at least one method in *REIGNS*' class and one in *MARRIAGES*' class; moreover, as it also contains functions defined on *RULERS* (i.e., *Sex*, *Father*, and *Mother*), you also need at least one method in the *RULERS*' class.

d) You must track value changes for any column corresponding to the functions occurring in the corresponding constraint formula. Keep in mind that almost any value update in any of these columns might violate the corresponding constraint. Dually, however, except valid changes from checking (e.g., when users delete one or both parents of a person corresponding constraints might not be affected, whereas replacing null with non-null values generally is potentially harmful). For example, to enforce $C_{26}$, you need to track changes in columns *Country*, *FromY*, *ToY*, and *Ruler* from table *REIGNS*, *Sex*, *Father*, and *Mother* from *RULERS*, as well as *Husband* and *Wife* from *MARRIAGES*. In this case, for example, if the current *RULERS* line contains 'N' in the *Sex* column, you can stop checking values in all other columns.

```
-- —— C6: Age ∈ [0,140] for Sex ≠ 'N' ———
IF EXISTS (
    SELECT 1 FROM inserted i
    WHERE  i.Sex <> 'N'
      AND  i.Age IS NOT NULL
      AND  (i.Age < 0 OR i.Age > 140))
BEGIN
    RAISERROR('Anybody''s age must be a ' +
    'natural at most equal to 140.', 16, 1);
    ROLLBACK TRANSACTION; RETURN;
END
```

Figure 20: T-SQL code generated by *MatBase* for enforcing constraint $C_6$

e) If only one column of a table is involved in a constraint, you should always use the corresponding VBA *column_BeforeUpdate* (.Net *Validating*) method-type, like in Fig. 16 for $C_2$: this way, users are immediately told if their new value is invalid and must fix it on the spot.

f) If more than one column of a table is involved in a constraint, you should always use the corresponding VBA *Form_BeforeUpdate* (.Net *Validating*) method-type, like in Fig. 10 for $C_6$: otherwise, you implicitly impose users an order in which to make the updates and sometimes prevent them from entering valid data.

g) Do not track the keys typed by users: let them finish typing and start enforcing constraints only after they (implicitly or explicitly) ask to save the data they just entered. Consequently, for example, never use VBA methods of the type *KeyPress*, *KeyUp*, *KeyDown*, *Change*, etc. for enforcing constraints. Tracking key-typing must be used only when db software applications need to provide hot keys of their own.

h) Do not ever-present users in combo-boxes with invalid data, but only with valid one. Use the VBA *Form_Current* method-type for filtering valid data, like in Fig. 14.

i) Do not hesitate to use even VBA *column/Form_AfterUpdate* (.Net *Validated*) methods for enforcing constraints. For example, as mentioned above, when enforcing constraint $C_4$ (the founder of any dynasty must belong to that dynasty), we should select in the *Founder* combo-box of form *DYNASTIES* all persons from *RULERS* belonging to the current dynasty or not belonging to any one; whenever users choose a person not belonging to a dynasty, we must use the *Form_AfterUpdate* of this class to run a SQL UPDATE query to set his/her dynasty to the current one.

j) Similarly, also use the VBA *Form_Delete* event method to enforce constraints containing a variable quantified existentially. For example, constraint $C_{26}$ might be also violated by deleting data of a marriage between two persons not related as parent – child, whenever they both simultaneously reigned over a same country; consequently, such deletion must be rejected.
k) Whenever possible, use (static or/and dynamic) blocking update/write access to columns or/and entire forms instead of writing complex code to achieve it (e.g., see Fig. 18).
l) For optimality reasons, do not ever check constraints for lines that, by definition, may not violate them. For example, when adding a new line of data in table *RULERS*, that person may not yet be the mother, father, or killer (not even for suicides!), or husband, or wife, or ruler, or dynasty founder; similarly, when adding a new line of data in table *CITIES*, that city may not yet be a country capital, etc.
m) Do not ever forget that, except for one-to-one (injective) ones, functions may take same values for several argument ones (i.e., are "many-to-one"/"one-to-many" relationships): consequently, you should check for only one value in codomain classes, but for several ones in the corresponding domain ones. For example, when enforcing $C_5$ (nobody may belong to a dynasty that was founded after his/her death), in the *RULERS* class you have at most to compare only the passed away year of the current person with the birth year of the founder of the corresponding dynasty (if any); dually, in the *DYNASTIES* class, you have to compare the birth year of the founder (if any) with the passed away year of all the persons from *RULERS* who belong to the current dynasty.
n) Generally, there is no need to duplicate enforcing the relational constraints with Windows class methods: all commercial RDBMSes are correctly enforcing them. However, sometimes you must do it, because, generally, interpreters first run your event-driven associated methods and only then, if they did not reject the corresponding updates, are checking the relational ones. For example, although both MS Access and SQL Server correctly enforce NOT NULL constraints, because $C_{20}$ needs to be enforced in the *MarriageYear_BeforeUpdate*, $C_{21}$ in the *DivorceYear_BeforeUpdate*, and $C_{30}$ in the *Form_BeforeUpdate* methods of class *MARRIAGES*, although both *Husband* and *Wife* are declared as NOT NULL in the table's scheme, if users enter data on a new line only in *MarriageYear* and/or *DivorceYear* and then move the cursor to the next new line, as both these methods need the corresponding *Husband* and *Wife* values, the application crashes if they do not first enforce their totality.
o) Always help users correct their mistakes as much as possible, not only by displaying context-dependent error messages indicating exactly the violated constraint and how to satisfy it (e.g., Fig. 17) but also restoring previous valid values (generally mistakenly modified).
p) Always replace context-independent system error messages with context-dependent ones: for example, even the least inconvenient ones of type "Are you sure you want to delete this line?" become much more user-friendly if rephrased as, e.g., "Are you sure you want to delete data for the Mar-a-Lago city from the Florida state of the U.S.A.?"; moreover, even if you do not have time or you are not paid for such "sweeties", replace at least the awful RDBMS error messages issued when users try to duplicate values in unique (sets of) columns: for example, a message of the type "A dynasty named Windsor is already stored in this table! Please give a unique name to the dynasty you want to add or cancel your request." will always make our customers happier.
q) Pay maximum attention to the treatment of null values, as this is the most complicated and tricky part of computer programming. First, as they may have several meanings: for example, the ones in *MarriageYear*, *BirthYear*, etc. means (temporarily) unknown values, while those in *PassedAwayYear*, *DivorceYear*, and *ToY*, may mean "unknown", but also still alive, married, ruling, respectively. Moreover, beware that treatment of null values is not standardized across interpreters and programming libraries.

## 8. CONCLUSIONS

We introduced and discussed the *M-R* (*A*7) *MatBase* pseudocode algorithm for translating our (E)MDM db schemes into relational ones and associated sets of non-relational constraints. We proved that this algorithm is very fast, sound, complete, and optimal. Of course, *M_R* may be also manually used by any db and software architect who does not have access to *MatBase*.

We applied this algorithm to a beautiful example from the genealogical trees subuniverse, made of 7 entity-type sets, 38 functions, and 119 constraints (82 relational and 37 non-relational).

We identified the implications between some of the 119 db constraints governing this subuniverse and eliminated 15 non-relational and 5 relational redundant ones.

We provided examples of both SQL and VBA code for enforcing non-relational constraints of all major types.

We also offered main 17 guidelines for designing and developing elegant and efficient code for enforcing non-relational constraints in user-friendly db software applications within our db constraint-driven framework.

The results of this research also prove once more the power and adequacy of our intelligent knowledge and database management system prototype *MatBase*.

Further work will be done to investigate enforcing non-relational constraints with the help of Artificial Intelligence tools like Claude Code, GitHub, Cursor, Windsurf, Replit, etc.